# A Modular Origami-inspired Mechanical Metamaterial

*Y. Yang, Z. You*

**Abstract:** Mechanical metamaterials with complex microstructures have superior physical properties such as graded stiffness, negative Poisson's ratio, and advantage in energy absorption. In recent years, origami provide many inspirations in the geometry structure of the metamaterials. Here we present a modular origami inspired reconfigurable metamaterial. Our approach exploits a transformable module consisted of the Sarrus linkage and planar four-bar linkage, and the modules are connected following a tessellation pattern to form a periodic structure. Different arrangements, module shapes and grid pattern of the metamaterial are discussed, and the loading capacity and stiffness variation are analysed experimentally. The proposed material can be used to realize reconfigurable structures or architectures over a wide range of length scales.

## 1 Introduction

Appropriately designed cellular micro-structures in metamaterials often result in superior mechanical properties, such as light weight, low thermal conductivity and negative Poisson's ratio. Recent years, transformable origami structures have become the source of inspiration in the unit cell design as the internal mechanisms of origami allow various configurations during folding, each of which could correspond to a different mechanical behaviour. For instance, it has been found that in a widely used origami pattern, the Miura-ori, changing the crease angles can lead to locking and pop-through transformations, which in turn, alters the overall stiffness of the material made from such a pattern [Schenk and Guest 13, Silverberg et al. 14, Yasuda and Yang 15, Filipov, et al. 15, Tachi and Miura 12, Cheung, et al. 14, Li, et al. 16]. However, most of the bistable features and local transformation designs presented so far are limited to surfaces made from a single sheet. Though stacking such sheets is possible, the variation of the shapes of the packed material is largely restricted because of the geometrical compatibility requirement of the single degree-of-freedom (DOF) structure layering. Therefore, finding a truly spatial assembly, instead of simple stacking of sheets, is probably a more suitable approach for the construction of metamaterial.

In this paper, the attention is drawn upon designing metamaterials by combining modular origami and spatial linkages. Modular origami is an art that uses a certain number of paper-folded units to assemble a structure [Gurkewitz and Bennett 12]. The authors have shown that concepts of modular origami could be adopted and extended to create single layered (two-dimensional) assemblies with particular properties using planar $4R$ linkages and their assemblies, where $R$ represents revolute hinges [Yang and You 18]. To take the single layer structure into a spatial assembly whilst keeping the motion features, proper transitional mechanism must be utilised. Here we adopt the well-known Sarrus linkage for this purpose. The Sarrus linkage is the first published overconstrained mechanism by Pierre Frédéric Sarrus in 1853 [Bennett 1905]. It consists of six links connected by two sets of parallel hinges forming a loop, and is capable of rectilinear motion where the top link can move vertically up and down to the base link. Kinematically it is known as a $6R$ spatial linkage. Using the Sarrus linkage as corner units, the two-dimensional modular origami can be transformed into a three-dimensional assembly, leading to various novel transformable cellular micro-structures.

The layout of this paper is as follows. The construction of modular origami building blocks, known as the Sarrus modular origami (SMO) modules, is first introduced, and it can be duplicated according to tessellation patterns to form a material grid. We demonstrate that a series of SMO modules can be obtained by combining planar $4R$ linkages and the Sarrus linkages. The modules use similar mechanism components, but go through different shape variation due to the geometrical discrepancy. Then the mechanical properties of some of the modules are analysed, including the theoretical evaluation of the Poisson's ratio and mechanical experiments.

## 2  Structure Design

### 2.1  Design of the Sarrus Modular Origami (SMO) module

A Sarrus linkage is shown in Figure 1(a). It has six identical cubes connected by two sets of hinges: each set consists of three parallel hinges and the hinge angle between these two sets of hinges must be neither 0 nor 180°. It has one DOF so top cube A is able to move vertically up and down with respect to bottom cube B. Hereafter cubes A and B that only have translation motions are named corner cubes, and the cube R, S, T and U which have both rotation and translation motions are called connector cubes.

Now connecting two Sarrus linkages with a planar $4R$ linkage made from four cubes of same dimension, we obtain an assembly shown in Figure 1(b) in which two Sarrus linkages are corners and the planar $4R$ linkage forms the side wall. The planar $4R$ linkage and the Sarrus linkage share the same cubes R, S, T and U. Should there be four corners and four walls with the hinge angle of each Sarrus linkage being 90°, we obtain a basic SMO module that has one DOF. Figure 1(c-e) shows transformation sequence of the module, and it has two limit states when the distance between cubes A and B reaches the minimum and the maximum. Looking from the top view, the square profile of the SMO module remains unchanged during the transformation.

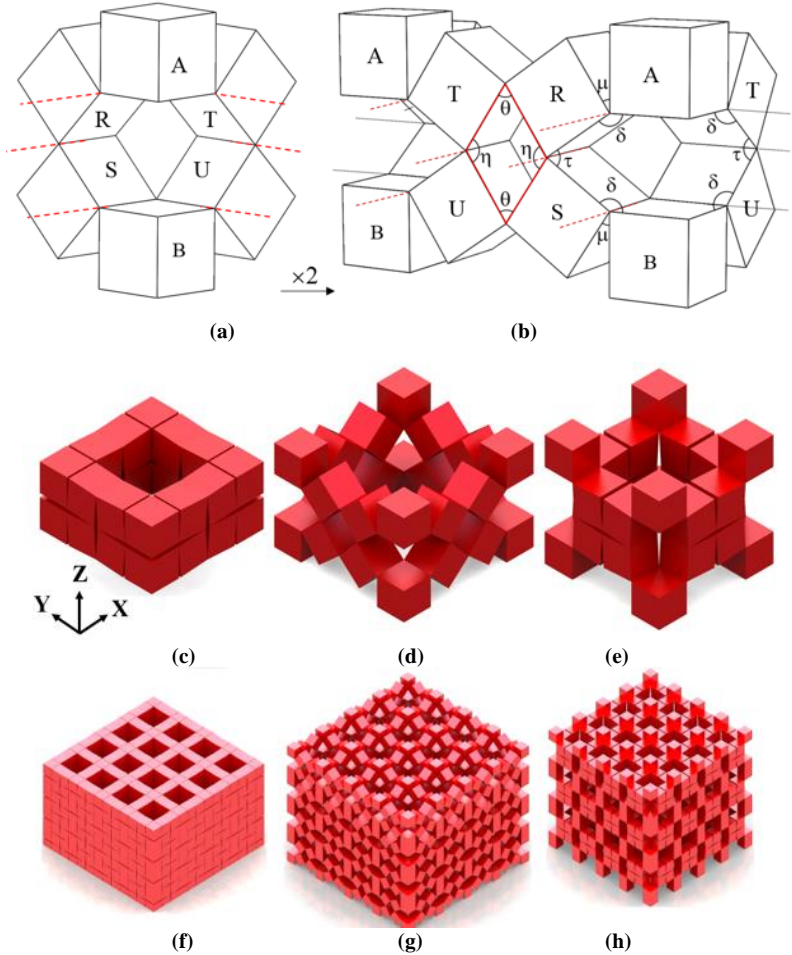

**Figure 1:** *Geometry design of the SMO module. (a) A Sarrus mechanism by cube units. (b) Connection of Sarrus mechanisms. (c-e) Transformation of a SMO module. (f-h) Deformation of a SMO metamaterial.*

Take $\theta$ shown in Figure 1(b) as the kinematic variable. If all of the cubes are the same, there are

$$\eta = \pi - \theta \tag{2.1}$$

$$\tau = \theta \tag{2.2}$$

$$\delta = \pi - \frac{\theta}{2} \tag{2.3}$$

$$\mu = \frac{\theta}{2} \tag{2.4}$$

Hence, it is clear that every angle can be uniquely determined by $\theta$. That is to say, the module has a single DOF.

Duplicating the SMO module in three orthogonal directions results in the structure for a metamaterial. Its configurations are shown in Figure 1(f-h). Originally, it is a porous structure where all the pores have the same orientation. During the transformation, the dimensions of the overall structure increases in three orthogonal directions, which indicates that as a material, it will have a negative Poisson's ratio. In the final state, the material ends up with pores in all three directions and the top and bottom surfaces are uneven. In short, the material built from the SMO modules will have different stiffness, density and porous orientation at different configurations.

A 3D printed SMO metamaterial sample is shown in Figure 2. Due to the difficulty in fabricating transformable porous materials with high complexity by the existing additive manufacturing method, 27 SMO modules were printed using a Stratasys Objet 500 Connex3 3D printer individually before they were assembled together. The cube size was 7×7×7mm. Slots were made on corner cubes to help align the modules, and a fast curing adhesive was used to permanently fix the modules together.

The behaviour of the prototype validates the geometric transformation.

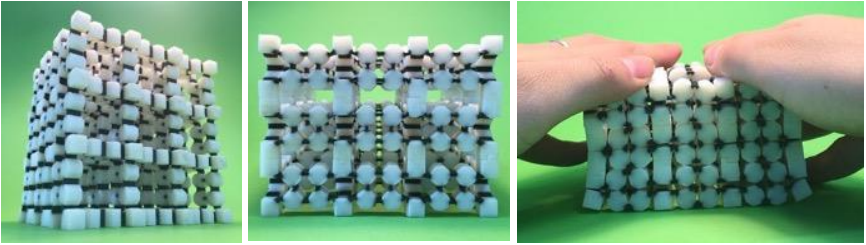

**Figure 2:** *3D printed SMO metamaterial and its geometric transformation.*

## 2.2  Planar 4*R* linkage assembly design in SMO modules

The basic SMO module illustrated above can be generalized by varying the shape and connection order of the planar 4*R* linkages and the Sarrus linkages. The single 4*R* linkage on each side can be replaced by other planar assemblies of 4*R* linkages outlined in [Yang and You 18]. Figure 3(a) shows two variations of modules where the original planar 4*R* linkage with 2×2 cubes is replaced by planar 4*R* linkage assemblies with 3×3 and 4×4 cubes, respectively. It can be seen that when the number of cubes in the planar assembly is even, the module forms a square tube when closed, and the upper and bottom surfaces are flat. If, however, the number of cubes in the planar assembly is odd, the closed state will have uneven top and bottom surfaces for the corner cubes pop up or down symmetrically. Figure 3(b) shows a piece of SMO metamaterial that constructed by the 4×4 SMO modules.

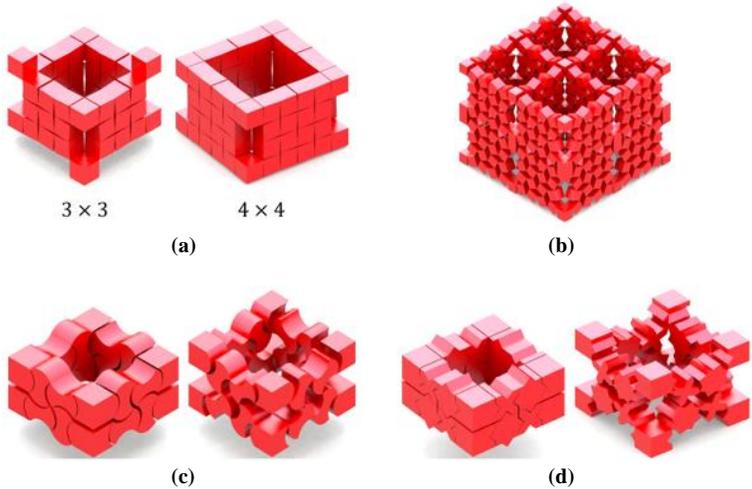

**Figure 3:** *4R linkage or linkage assembly design. (a) SMO modules whose sides have more cubes. (b) SMO metamaterial by 4×4 SMO modules. (c) SMO module where cubes are replaced by solids with curved edges. (d) SMO module with paired solids.*

It is also possible to replace the connector cubes in the planar $4R$ assemblies with solids other than the cube, such as concave polygons or cells with curved edges as in Figure 3(c), and to do so, the corner cubes must be adjusted accordingly. For instance, they could be made the same as the solids in the planar $4R$ assemblies. Moreover, paired shapes can be applied to the connectors as well. Figure 3(d) shows an SMO module with jigsaw pair shapes. The advantage of using solids with curve edges or pair shapes is that they have better shearing resistance when packed because the customised shapes provide larger contact surfaces between neighbouring solids.

All of the variations of solid shapes outlined in [Yang and You 18] can be used to construct the planar $4R$ assemblies.

## 2.3 Sarrus linkage design of SMO modules

The top views of SMO modules presented are planar diagrams consisting of squares and voids (Figure 4(a)). These shapes can be changed using tiling technique [Grünbaum and Shephard 87]. For instance, Figure 4(b) shows a tiling pattern made from triangles, squares and hexagons. The triangles (shown in grey) can be taken as corner cubes of the Sarrus linkages whereas the squares form the side walls and hexagons are voids. As a result, the hinge angle of each Sarrus linkage becomes 120°. Figure 4(c) is another tiling pattern made from triangles, squares and hexagons, and in this case hexagons are taken as corner cubes and triangles are voids. The hinge angle becomes 60°. Figure 4(d) is a third case that adopt both hexagons and triangles as corner cubes of Sarrus linkage. Following this rule, one can design the grid of SMO metamaterial using various uniform tilings. The shape of the corner cube determines the hinge angle of the Sarrus linkage in SMO module. Then, the modules can be connected following the tessellation patterns to form metamaterials of different geometry as in Figure 4(e-g).

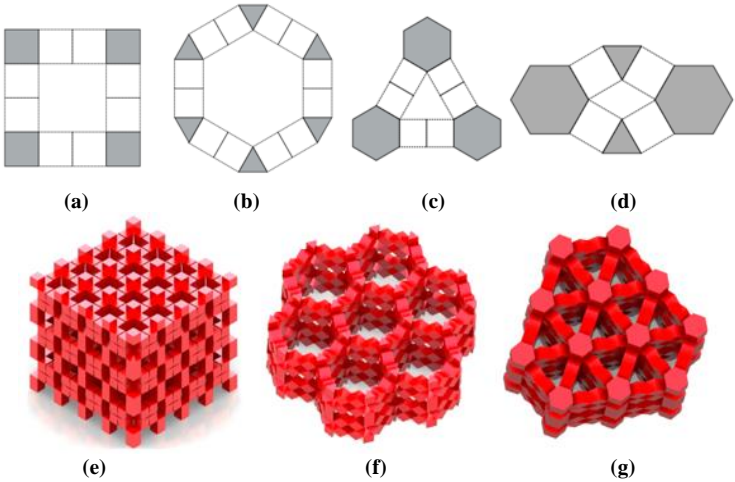

**Figure 4:** *Sarrus linkage design. (a-d) Corner cube shape variation. (e-g) SMO metamaterials in different tiling patterns.*

## 3  Mechanical Properties

### 3.1  The Poisson's ratio

The Poisson's ratios are calculated to evaluate the dimensional and volumetric change of the SMO modules. Assuming that the solids in SMO modules are rigid, i.e., they can neither bend nor stretch, and the width of hinges is ignored, the motion of modules is fully determined by kinematics. Therefore, the Poisson's ratios and engineering strains can be theoretically calculated from geometrical quantities alone.

Consider a basic SMO module shown in Figure 1(c). Let $a$ be the edge length of the cubes, and change the moving angle $\theta$ from 0 to 180° in Figure 1(b). If the module is loaded along the vertical direction ($z$), the width $w$ (–x axis), depth $d$ (–y axis) and height $h$ of the module can be calculated as

$$w = d = 2a(1+\sin\frac{\theta}{2}+\cos\frac{\theta}{2}) \qquad (3.1)$$

$$h = 2a(1+\sin\frac{\theta}{2}) \qquad (3.2)$$

Initially when the SMO module is close packed (Figure 1(c)), the edge length $w_0$, $d_0$ and $h_0$ are given

$$w_0 = d_0 = 4a, h_0 = 2a \qquad (3.3)$$

The variation of the edge length during transformation can be calculated

$$dw = w - w_0 = 2a(\sin\frac{\theta}{2}+\cos\frac{\theta}{2}-1) \qquad (3.4)$$

$$dh = h - h_0 = 2a\sin\frac{\theta}{2} \qquad (3.5)$$

Accordingly, the axial strain $d\varepsilon_z$, transverse strains $d\varepsilon_x$ and $d\varepsilon_y$, and Poisson's ratio $v_x$ (–x axis) and $v_y$ (–y axis) are derived

$$d\varepsilon_z = \frac{dh}{h_0} = \sin\frac{\theta}{2}$$

$$d\varepsilon_x = \frac{dw}{w_0} = \frac{\sin\frac{\theta}{2}+\cos\frac{\theta}{2}-1}{2} = d\varepsilon_y \qquad (3.6)$$

$$v_x = -\frac{d\varepsilon_{trans}}{d\varepsilon_{axial}} = -\frac{d\varepsilon_x}{d\varepsilon_z} = -\frac{h_0 \cdot dw}{w_0 \cdot dh} = \frac{(\sin\frac{\theta}{2}+\cos\frac{\theta}{2}-1)}{2\sin\frac{\theta}{2}} = v_y \qquad (3.7)$$

For a metamaterial with $n \times n \times n$ such SMO modules, the material dimensions are $w_n$, $d_n$ and $h_n$, and original dimensions $w_{n0}$, $d_{n0}$ and $h_{n0}$ are as follows

$$w_n = d_n = (n+1)a + 2na(\sin\frac{\theta}{2}+\cos\frac{\theta}{2}) \qquad (3.8)$$

$$h_n = 2na(1+\sin\frac{\theta}{2}) \qquad (3.9)$$

$$w_{n0} = d_{n0} = (3n+1)a, \ h_{n0} = 2na \qquad (3.10)$$

The Poisson's ratio $v_{nx}$ (–x axis) is derived

$$v_{nx} = -\frac{h_{n0} \cdot dw}{w_{n0} \cdot dh} = -\frac{2n(\sin\frac{\theta}{2}+\cos\frac{\theta}{2}-1)}{(3n+1)\sin\frac{\theta}{2}} \qquad (3.11)$$

The Poisson's ratios of SMO module and metamaterials are plotted in Figure 5. For the SMO module in Figure 5(a), the Poisson's ratio is negative throughout the opening process, for the lengths along x- and y-axes exceed the original length with the stretch in z-direction. It is also noted that, with the increase in layer amount in a SMO metamaterial, the Poisson's ratio will vary accordingly due to the corner cube number. The comparison of Poisson's ratio with different layer amounts is shown in Figure 5(b). This feature can be used to design metamaterials with various volume change requirement.

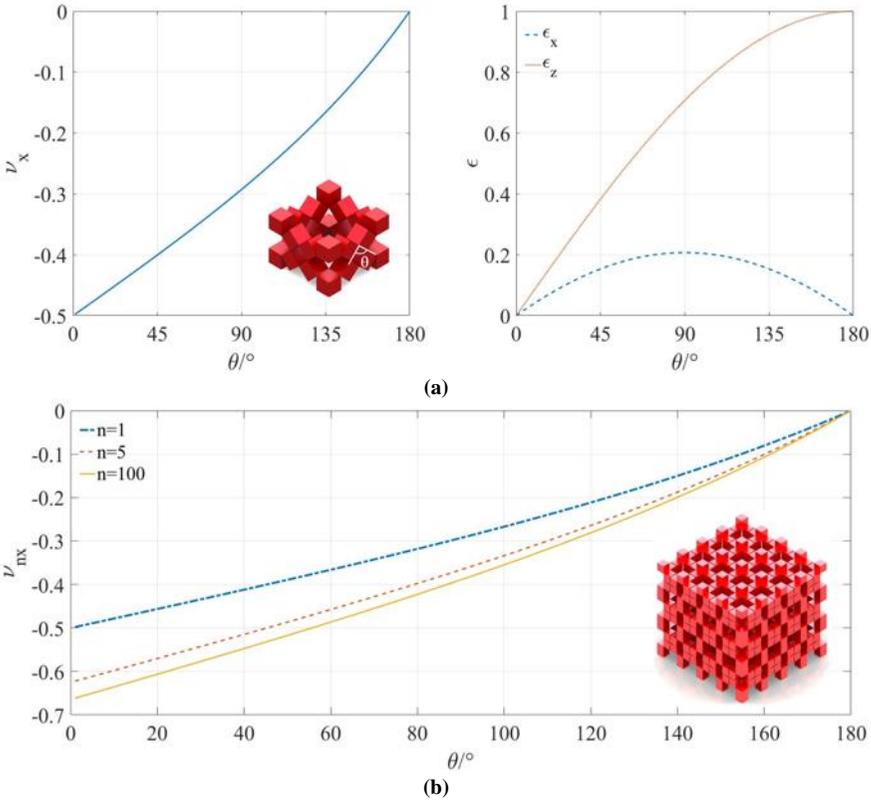

**Figure 5:** *Poisson's ratio. (a)Poisson's ratio and engineering strain of a single SMO module. (b)Poisson's ratio of multi-module metamaterials.*

## 3.2 Manufacturing and mechanics experiments

The prototype used to validate motion of the metamaterial, Figure 2, consisted relatively rigid cubes and elastic hinges. Under vertical axial compression (along $z$ direction), a two stage stiffness variation can be obtained. It is expected that under external loading, the material would first exhibit lower stiffness because the deformation is induced by relative rotations amongst cubes, and then the stiffness increases after all cubes pack together and they start deforming. Moreover, the stiffness at the second stage can be adjusted if the cubes are made less rigid materials.

To verify this, two single module test samples were made based on the SMO module shown in Figure 1(c). Both samples were made of multi Polyjet photopolymers material: the first one use rigid material for cubic solids and a rubber-like flexible material for hinges, and the second use the rubbery material for whole part. The solid cubes in the first sample have a side length of 10mm, whereas the second hollow cubes of same dimension with channels of square section. The directions of channels are perpendicular to the loading direction and the wall thickness was 2mm. The hollow cubes can generate further deformation under compression in the close pack configuration of the structure.

The response of the SMO module is tested by uniaxial compression using an Instron with a 100N force at a compression rate of 0.16mm/s. The samples were loaded until all cubes were completely closed. For the hollow cube SMO module, a distance of 5mm was further pressed after complete closure of the sample.

The compression test results are shown in Figure 6. The results indicate that for the solid cube SMO module, the load required to cause mechanism motion was very low, and the stiffness (gradient of the force-displacement curve) was relatively low. After the cubes were close packed, the load required to generate further material deformation increased sharply. For the hollow cube module, the force-displacement curve displayed the same feature during the mechanism motion: when it was nearer to the fully packed state, the stiffness started to increase, indicating that the transformation of SMO module began to take place. This was followed by a sharp increase of stiffness, which is lower than the solid cube sample because the cubes are hollow. The force then reached a peak at which some of the hollow cubes buckled. When this secondary deformation ceases, the load increases again because of the real material compression.

These experiments have proved that the metamaterials made from SMO modules can have programmable stiffness, and the stiffness and peak load can be adjusted by adopting specific cubes. Note that the observed behaviour is not fully elastic, and the load capacity is highly influenced by the stiffness and resistance of the hinge material.

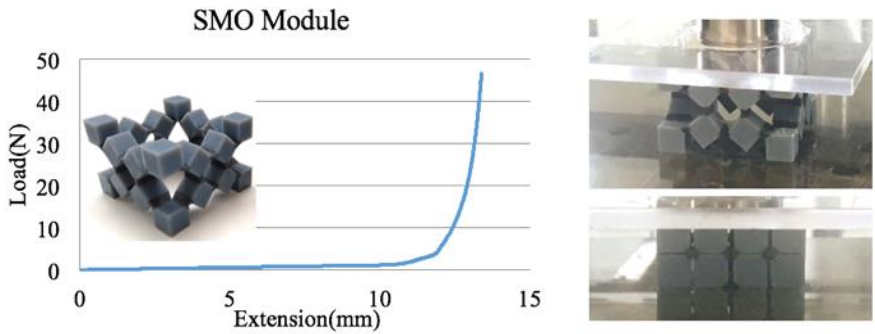

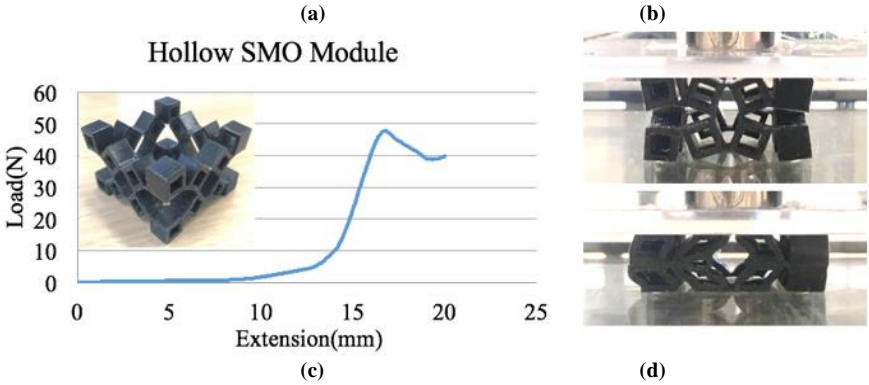

**Figure 6:** *Compression tests of SMO modules. (a) Force-deformation curve of solid SMO module. (b) Deformation of a solid SMO module. (c) Force-deformation curve of hollow SMO module. (d) Deformation of a hollow SMO module.*

## 4 Conclusions

In this work, a programmable three-dimensional mechanical metamaterial has been introduced inspired by modular origami. The Sarrus linkage is adopted to create a truly three-dimensional structure known as the SMO module, and the tessellation of the SMO modules leads to the framework of the metamaterial. A family of structures with different geometrical features are proposed resulting in various configurations. 3D printed samples have been fabricated and tested to study the mechanical properties of the material. Under external loading, the material first undergoes kinematic motion, followed by structural deformation after all the modules are packed together. The programmable stiffness feature of this material makes it a suitable material to construct body armours and automotive structures [Valdevit et al. 11, Schaedler et al. 14].

The construction strategy in this study of linking spatial linkages following a tessellation pattern can be used to design other metamaterials. For instance, the Bennett linkage and Myard linkage have been used to build motion structures [Chen and You 05, Liu and Chen 09]. Moreover, the solids in the morphing module can be replaced by tubular origami structures, e.g., the origami crash box [Ma and You 11], to obtain tunable stiffness and better energy absorption capacity.

Because of the scale-free geometric character of origami, the designs explored in this paper can be realized in micro- and nano-scale or extended to the meter-scale to build deployable domes and transformable architecture material. In the past, 3D microstructure manufacturing methods such as deep UV, LIGA and projection micro-stereolithography have been used to fabricate microlattices [Sun et al. 05, Jang, et al. 13]. These techniques may also be used to produce metamaterials proposed here. Moreover, the modules introduced may also be applied to create self-assembly and laminate-based mechanisms [Aukes et al. 14, Felton, et al. 15]. In conclusion, this research provides food for thought not only in mechanical metamaterial, but also in design of motion structures over a range of dimensional scales.

## Acknowledgements

The authors wish to acknowledge the support of Air Force Office of Scientific Research (FA9550-16-1-0339). Yang would like to thank the financial support from the Clarendon Scholarship.